\documentclass[english, 11pt, a4paper]{article}

\usepackage{slashed}
\usepackage[utf8]{inputenc}
\usepackage{babel}
\usepackage{graphics}
\usepackage{graphicx}
\usepackage{texdraw}
\usepackage{epsfig}
\usepackage{float}
\usepackage{amsmath}
\usepackage{amssymb}
\usepackage{enumerate}
\usepackage[all]{xy}
\usepackage{fullpage}
\usepackage[small]{caption}
\usepackage{hyperref}
\usepackage{physics}
\usepackage{cite}

\newcommand{\be}{\begin{equation}}
\newcommand{\ee}{\end{equation}}

\providecommand{\abs}[1]{\lvert#1\rvert}

\begin{document}

\title{\bf Spectral density of the Dirac-Ginsparg-Wilson operator, chiral $U(1)_A$ anomaly, and analyticity in the high temperature
  phase of $QCD$}

\date{}
\author{Vicente Azcoiti \\
        Departamento de F{\'i}sica Te\'orica, Facultad de Ciencias, and \\
        Centro de Astropartículas y Física de Altas Energías (CAPA),\\ Universidad de Zaragoza,
        Pedro Cerbuna 9, 50009 Zaragoza, Spain\\}

\maketitle

\begin{abstract}
  Using general properties of the $Q=0$ topological sector we previously argued that
  a vectorlike theory, with chiral $U(1)_A$ anomaly, and exact non-Abelian chiral symmetry, should exhibit divergent
  susceptibilities in the chiral limit,
the two-flavor Schwinger model being a paradigmatic example of the realization of this scenario. Two flavor 
$QCD$ at $T>T_c$ satisfies all the above conditions, and it is also expected that the $U(1)_A$ axial symmetry remains effectively
broken in its high temperature phase. Therefore we would expect a nonanalyticity in the quark mass
dependence of the free energy density, in contrast with the Dilute Instanton Gas Approximation (DIGA) prediction.
We investigate in this work whether the aforementioned results can also be reproduced making only use of standard
properties of the spectral density of the Dirac operator, without having to resort to general properties of the $Q=0$
topological sector. We show that the only way to derive a nontrivial $\theta$-dependence, and an analytical free energy
density in $QCD$ with two degenerate flavors is that the spectral density, $\rho\left(\lambda,m\right)$, of the
absolute value of the nonzero modes of the Dirac-Ginsparg-Wilson operator develops a $m^2\delta(\lambda)$
function in the thermodynamic limit. This is the expected result in the DIGA, where interactions between instantons
in the dilute gas are fully neglected. However, 
at temperatures close to $T_c$
the interaction between instantons should become non-negligible, and
the splitting from zero of the near-zero modes, which has been neglected in the DIGA, 
should be taken into account. Therefore we expect that the
$m^2\delta(\lambda)$ contribution to the spectral density is no longer correct at these temperatures, and that the free energy
density becomes a nonanalytic function of the quark mass.

\end{abstract}

\vfill\eject

\section{Introduction} \label{intro}

Understanding the fate of the axial $U(1)_A$ anomaly in the high temperature
phase of $QCD$ is a great challenge for high energy theorists. Indeed the axion,
predicted by Weinberg \cite{weinberg} and Wilczek \cite{wilczek}, in the Peccei and Quinn mechanism \cite{pq},
is one of the more interesting candidates to make the dark matter of the universe, and the axion potential,
which determines the dynamics of the axion field, is closely linked to the topological properties of QCD, and
in particular to the topological susceptibility. Moreover it has been argued that scalar and
pseudoscalar meson screening masses, and their corresponding susceptibilities, are very sensitive to the
realization of the $U(1)_A$ axial anomaly in the high-temperature chiral symmetric phase of $QCD$. 
More precisely it has been argued that if the $U(1)_A$ symmetry remains
effectively broken i.e., order parameters for the $U(1)_A$ symmetry take nonvanishing vacuum expectation values
in the chiral limit when compatible with the realization of other symmetries, 
the topological properties of the theory can be the basis of a new mechanism, other
than Goldstone’s theorem, to generate a rich spectrum of quasimassless bosons near the chiral limit
\cite{vic1,vic2,vic3}.

Computing the topological susceptibility in the high temperature phase of QCD from first principles,
through numerical simulations in the lattice, is a very hard task because of several
nontrivial numerical problems \cite{guido}. First, there is a sampling problem because the probability of the
vanishing topological charge sector, $Q=0$, is much higher than the probability of any $Q\ne 0$ sector
on affordable volumes. Second, the numerical determination of the topological susceptibility is
affected by large lattice artifacts because the lattice fermions used explicitly break the chiral symmetry.
Moreover, for small enough values of the lattice spacing, the topological critical slowing down phenomenon
is present in local updating algorithms, and ergodicity is lost. We refer the interested reader to Ref. 
\cite{guido} to get an idea of the current situation on this topic.

On the other hand, given these technical difficulties, we can also resort to some simplified models, as the
Dilute Instanton Gas Approximation, which is expected to work at asymptotically high temperatures.
The DIGA \cite{DIGA} assumes that at temperatures much higher than $T_c$, Debye  
screening would only allow instantons of very small radius to exist, and hence the $QCD$ vacuum  
energy density of noninteracting instantons should not suffer from infrared singularities. The  
vacuum energy density can therefore be expanded in powers of the quark masses, and the
topological susceptibility $\chi_T$ of $QCD$ with $N_f$ quark flavors behaves like 

\begin{equation}
 \chi_T \approx C\left(T\right)\prod_{i=1}^{N_f} m_i.
  \label{uno}
\end{equation}

Lattice simulations of pure gauge $SU(2)$ and $SU(3)$ gauge theories have confirmed the DIGA prediction at $T\sim 2T_c$ \cite{urs}, where $T_c$ is the deconfinement temperature.
However
in $QCD$, with two light flavors, instanton contributions to the partition function are very suppressed by the fermion
determinant, and the validity of the DIGA, in what concerns the light quark mass dependence of the
topological susceptibility (\ref{uno}), has not been tested. Rather it has been used, for instance in Ref. \cite{petre},
to re-scale their determinations of $\chi_T$, obtained at a Goldstone pion mass of $160 MeV$, to the physical point.

Using general properties of the $Q=0$ topological sector \cite{vic0} we argued in
Refs. \cite{vic1,vic2,vic3} that 
a gauge-fermion quantum field theory, in which the $U(1)_A$ axial symmetry remains effectively broken, and 
where the chiral condensate vanishes in the chiral limit, because of a not spontaneously broken
non-Abelian chiral symmetry,
should exhibit a divergent correlation length in the correlation function of the scalar condensate in
the chiral limit. In such a case also some pseudoscalar correlation functions, associated to what would be the
Nambu-Goldstone bosons if the non-Abelian chiral symmetry were spontaneously broken,
should exhibit a divergent correlation length. 
These results, if applied to the high-temperature phase of $QCD$, would imply a nonanalytical behavior
of the free energy density in the quark masses, in contrast to the DIGA prediction (\ref{uno}).

The main goal of this work is to investigate whether the aforementioned results can also be reproduced 
making only use of standard properties of the spectral density of the Dirac operator, 
without having to resort to general properties of the $Q=0$ topological sector. With this purpose we will present 
our theoretical setup, which makes use of lattice Ginsparg-Wilson fermions, in section \ref{spectral}.
We will also calculate in this section the scalar condensate in one-flavor $QCD$, with the help of the spectral
density of the Dirac-Ginsparg-Wilson operator, and will remember how the Banks-Casher mechanism \cite{b-c}
allows us to reproduce the expected result for the scalar condensate in the chiral limit.
In section \ref{spectraltwo}, which contains the main body of this work, we will
show that the only way to obtain an analytic free energy density with nontrivial topology, in the high temperature
chiral symmetric phase of two-flavor $QCD$, is that the spectral density,
$\rho(\lambda,m)$, 
of the absolute value of the nonzero modes of the Dirac-Ginsparg-Wilson operator, develops a $m^2\delta(\lambda)$ 
contribution in the thermodynamic limit. This is the expected result in the DIGA, 
where interactions between instantons are fully neglected, an approximation that may be reliable at very high temperatures.
However at temperatures higher but close to $T_c$, the critical temperature of the chiral $SU(2)_A$ restoration transition,
the interaction between instantons should become non-negligible. Therefore, the splitting from zero of the near-zero
modes, which
has been neglected when assuming the $\delta(\lambda)$ behavior in the spectral density $\rho(\lambda,m)$,
should be taken into account. We analyze this issue in section \ref{con}, which contains our conclusions.

\section{Spectral density of the Dirac-Ginsparg-Wilson operator: one-flavor case} \label{spectral}

The $QCD$ Euclidean continuum action with a $\theta$-vacuum term is 

\begin{equation}
  S = \int d^{4}x \left\{\sum^{N_f}_f\bar\psi_f\left(x\right)
  \left(\gamma_\mu D_\mu\left(x\right)+ m_f\right) \psi_f\left(x\right)
  + \frac{1}{4} F^a_{\mu\nu}\left(x\right)F^a_{\mu\nu}\left(x\right)
  + i\theta Q\left(x\right)\right\}
  \label{eulagran}
\end{equation}
where $D_\mu(x)$ is the covariant derivative, $N_f$ the
number of flavors, and $Q(x)$ the density of topological charge of the gauge configuration, whose integral
over the space-time volume is an integer number, the quantized topological charge $\nu$,

\begin{equation}
  \nu = \frac{g^2}{64\pi^2} \int d^4x\epsilon_{\mu\nu\rho\sigma}
  F^a_{\mu\nu}\left(x\right)F^a_{\rho\sigma}\left(x\right).
  \label{ftopcharg}
\end{equation}

To avoid ultraviolet divergences we will use a lattice
regularization, and Ginsparg-Wilson (G-W) fermions \cite{Ginsparg},
that share with the continuum formulation all essential ingredients: an explicit $U(1)_A$ anomalous
symmetry \cite{Luscher}, good chiral properties, a quantized topological charge, and an exact index
theorem on the lattice \cite{Victor}.

The G-W one-flavor fermion action can be written in a compact form as

$$S_F = a^4\left( \bar\psi D\psi + m\bar\psi \left(1-\frac{a}{2}D\right)\psi\right)=$$
\begin{equation}
a^4 \sum_{v, w} \bar\psi\left(v\right) D\left(v, w\right)\psi\left(w\right) +
  a^4 m \sum_{v, w} \bar\psi\left(v\right) \left(1-\frac{a}{2}D\left(v, w\right)\right)\psi\left(w\right)
  \label{fa}
\end{equation}
where $v$ and $w$ contain site, Dirac and color indices, and the G-W operator, $D$, is
$\gamma_5-Hermitian$

\begin{equation}
  \gamma_5D\gamma_5 = D^{\dagger}
  \label{hermi}
\end{equation}
and obeys the essential anticommutation relation

\begin{equation}
  D\gamma_5 + \gamma_5D = a D\gamma_5D
  \label{antic}
\end{equation}
$a$ being the lattice spacing.

Equations (\ref{hermi}) and (\ref{antic}) determine the main properties of the spectrum of the G-W
operator, which we can summarize as follows \cite{langlib}:

\begin{enumerate}
\item
  The eigenvalues $\mu$ of $D$ are either real or come in complex conjugate pairs.
  
\item
  Only eigenvectors with real eigenvalues can have nonvanishing chirality.

\item
  The eigenvalues $\mu$ are restricted to a circle in the complex plane, the Ginsparg–Wilson circle. 
  This circle has its center at $\frac{1}{a}$ on the real axis and a radius of $\frac{1}{a}$.
  \end{enumerate}

In the chiral limit, $m=0$, action (\ref{fa}) is invariant under the $U(1)_A$
chiral rotation

\begin{equation}
  \psi\rightarrow e^{i\alpha\gamma_5\left(I-\frac{1}{2}aD\right)}\psi,
  \hskip 1cm \bar\psi\rightarrow
\bar\psi e^{i\alpha\left(I-\frac{1}{2}aD\right)\gamma_5}
  \label{chirot}
\end{equation}
which for $a\rightarrow 0$ reduces to the standard continuum chiral transformation. 
However the integration measure is not invariant, and the change of
variables (\ref{chirot}) induces a Jacobian

\begin{equation}
e^{-i 2\alpha \frac{a}{2} tr\left(\gamma_5 D\right)}
  \label{jacobian}
\end{equation}
where

\begin{equation}
\frac{a}{2} tr\left(\gamma_5 D\right) = n_- - n_+ = \nu
  \label{topcharge}
\end{equation}
is an integer number, the difference between left-handed and right-handed zero modes, which is the
topological charge $\nu$ of the gauge configuration. Furthermore the scalar and pseudoscalar condensates

\begin{equation}
S = \bar\psi \left(1-\frac{a}{2}D\right)\psi \hskip 1cm 
P = i\bar\psi \gamma_5\left(1-\frac{a}{2}D\right)\psi
\label{sapc}
\end{equation}
transform under the chiral $U(1)_A$ rotations (\ref{chirot}) as a vector, just in the same
way as $\bar\psi\psi$ and $i\bar\psi\gamma_5\psi$ do in the continuum formulation.

The partition function $Z_\nu$ in the $\nu-topological$ sector can be written as follows

\begin{equation}
  Z_\nu = \left(2m\right)^{\abs{\nu}}
  \int_\nu \left[dU\right] \prod_j \left(m^2 +
  \left(1-\frac{1}{4}m^2\right)\lambda_j^2\left(U\right)\right) e^{-\beta S_G(U)}
\label{partitionnu}
\end{equation}
while for the full partition function we have

\begin{equation}
  Z = 
  \int \left[dU\right] \prod_j \left(m^2 + \left(1-\frac{1}{4}m^2\right)\lambda_j^2\left(U\right)\right)
  \left(2m\right)^{\abs{\nu\left(U\right)}} e^{-\beta S_G(U)}.
\label{partition}
\end{equation}
$S_G(U)$ is the pure gauge action, $m$ the quark mass in lattice units, $\beta$ the inverse gauge
coupling, $\nu(U)$ the topological charge of the gauge configuration, and

\begin{equation}
  \lambda_j\left(U\right) = \abs{\mu_j\left(U\right)}a
\label{eigenv}
\end{equation}
where $\mu_j(U)$ are the eigenvalues of the G-W operator $D$ which either come in complex conjugate
pairs, or are real with both chiralities, and the subscript $j$ runs over half the number of these
eigenvalues. The dimensionless quantity $\lambda_j$ takes values $0\le\lambda_j\le 2$.

The normalized scalar condensate is the logarithmic derivative of the partition function

\begin{equation}
  \left<s\right> = \frac{1}{V} \left<\bar\psi \left(1-\frac{a}{2}D\right)\psi\right> =
  -2m \left<\frac{1}{V}\sum_j\frac{1-\frac{1}{4}\lambda_j^2\left(U\right)}{m^2+
    \left(1-\frac{1}{4}m^2\right)\lambda_j^2\left(U\right)}\right> -
  \frac{1}{m} \left<\frac{\abs{\nu\left(U\right)}}{V}\right>
\label{scalarc0}
\end{equation}
where $V$ is the space-time volume in lattice units.

We can explicitly write the contribution to the condensate of all
exact zero modes, the unpaired $\abs{\nu\left(U\right)}$, plus occasional pairs of zero modes with opposite chiralities 
that may appear in the first addend of (\ref{scalarc0}), and proceeding in this way we get

\begin{equation}
  \left<s\right> = 
  -2m \left<\frac{1}{V}\sum^{n.z.}_j\frac{1-\frac{1}{4}\lambda_j^2\left(U\right)}{m^2+
    \left(1-\frac{1}{4}m^2\right)\lambda_j^2\left(U\right)}\right> -
    \frac{1}{m} \left<\frac{n_+\left(U\right)+n_-\left(U\right)}{V}\right>
\label{scalarc}
\end{equation}
and the sum runs only over half of the nonzero modes.

The standard wisdom on the vacuum structure of one-flavor $QCD$ 
in the chiral limit is that it is unique at each given value
of $\theta$. Indeed, the only plausible reason to
have a degenerate vacuum in the chiral limit would be the spontaneous
breakdown of chiral symmetry, but since it is anomalous, actually there is
no symmetry. Therefore the model is expected to show a mass gap in the chiral limit, and be free of infrared divergences.
The vacuum energy density can be expanded in powers of the fermion mass $m$, treating the quark mass term as a perturbation \cite{Smilga}, 
and this expansion will be then an ordinary Taylor series both, for the vacuum energy density

\begin{equation}
  E \left(m,\theta\right) = E_0
  - \Sigma m \cos\theta +O(m^2),
  \label{LS}
\end{equation}
and for the scalar condensate

\begin{equation}
  \left<s\right> =  -\Sigma\cos\theta +O(m).
\end{equation}

We will take in what follows $\theta=0$ and $m>0$. 
Since the thermodynamic and chiral limits commute, $\Sigma$ can be computed from
(\ref{scalarc}) as follows 

\begin{equation}
  \Sigma =
  \lim_{V\rightarrow\infty} \lim_{m\rightarrow 0}
  \frac{1}{m} \left<\frac{n_+\left(U\right)+n_-\left(U\right)}{V}\right> =
  \lim_{V\rightarrow\infty} \lim_{m\rightarrow 0}
  \left(\frac{2}{V m} \frac{Z_1}{Z_0}\right).
  \label{sigma1}
\end{equation}
which implies a finite density of zero-modes in the chiral limit that behaves as

$$
\lim_{m\rightarrow 0}\left<\frac{n_+\left(U\right)+n_-\left(U\right)}{V}\right>
  \approx m\Sigma.
$$

But we can also take the two limits in (\ref{scalarc}) in the opposite order, and in such a case we get

\begin{equation}
  \Sigma =
  \lim_{m\rightarrow 0} 2m\int_0^2 \frac{\rho\left(\lambda,m\right)\left(1-\frac{1}{4}\lambda^2\right)}
      {m^2+ \left(1-\frac{1}{4}m^2\right)\lambda^2}d\lambda +
      \lim_{m\rightarrow 0} \lim_{V\rightarrow\infty} \frac{1}{m}
      \left<\frac{n_+\left(U\right)+n_-\left(U\right)}{V}\right>.
  \label{sigma2}
\end{equation}
We have used in (\ref{sigma2}) the standard definition of the spectral density of the absolute value of the nonzero modes

\begin{equation}
  \rho\left(\lambda,m\right) = \lim_{V\rightarrow\infty}
  \left<\frac{1}{V}\sum^{n.z.}_j\delta\left(\lambda-\lambda_j\left(U\right)\right)\right>
  \label{spectrald}
\end{equation}
where $\lambda_j(U)$ is the absolute value of the paired complex-conjugate eigenvalues (\ref{eigenv}) 
of the Dirac-Ginsparg-Wilson operator.

Equation (\ref{sigma2}) can be satisfied in one of the following two ways:

\begin{itemize}
\item
The density of zero-modes vanishes in the infinite volume limit for any $m> 0$

\begin{equation}
\lim_{V\rightarrow\infty} \left<\frac{n_+\left(U\right)+n_-\left(U\right)}{V}\right>=0
\label{bankszm}
\end{equation}
and there is therefore no zero-mode contribution to the chiral condensate in the
thermodynamic limit. The scalar condensate, in this limit, is given by

  \begin{equation}
   \left<s\right> = -2m\int_0^2 \frac{\rho\left(\lambda,m\right)\left(1-\frac{1}{4}\lambda^2\right)}
  {m^2+ \left(1-\frac{1}{4}m^2\right)\lambda^2}d\lambda
\label{scalarc2}
\end{equation}
  where $\rho(\lambda,m)$ (\ref{spectrald}) is the density of the absolute value of the nonzero modes of the Dirac-Ginsparg-Wilson
  operator $D$.

\item
  The density of zero modes of the Dirac-Ginsparg-Wilson operator is finite in the infinite volume limit

  \begin{equation}
    \lim_{V\rightarrow\infty} \left<\frac{n_+\left(U\right)+n_-\left(U\right)}{V}\right> =
    m \Sigma + \dots
  \label{zeromodes}
\end{equation}
  and there is no contribution of the nonzero modes to the chiral condensate in the chiral limit.
\end{itemize}

The latter possibility has been excluded by the results of Refs. \cite{randommatrix1,randommatrix2}, where the
authors show that
zero modes do not contribute to the chiral condensate in the thermodynamic limit for $m>0$ and $\theta=0$.
Moreover, a 
finite density of zero modes in the infinite volume limit is quite implausible because in actual lattice calculations
one never finds zero modes of both chiralities. This means that $n_+(U)+n_-(U) = \abs{\nu(U)}$ and therefore

  \begin{equation}
    \lim_{V\rightarrow\infty} \left<\frac{n_+\left(U\right)+n_-\left(U\right)}{V}\right> =
    \lim_{V\rightarrow\infty} \left<\frac{\abs{\nu\left(U\right)}}{V}\right>
   \label{zeromodes2}
\end{equation}
  and the right-hand side of (\ref{zeromodes2}) vanishes for any non-negative value of the quark mass $m$ because otherwise parity would be spontaneously broken, and also because 
 the topological susceptibility is finite.

 We conclude that Eq. (\ref{sigma2}) is realized in the former of the two ways, and hence

  \begin{equation}
   \Sigma = \lim_{m\rightarrow 0} 2m\int_0^2 \frac{\rho\left(\lambda,m\right)\left(1-\frac{1}{4}\lambda^2\right)}
  {m^2+ \left(1-\frac{1}{4}m^2\right)\lambda^2}d\lambda.
\label{sigma3}
  \end{equation}
  This equation could be fulfilled if there is a singular contribution, $\Sigma m \delta(\lambda)$, to $\rho(\lambda,m)$. However 
  the standard wisdom is that equation (\ref{sigma3}) holds because the spectral density $\rho(\lambda,m)$
  remains finite for $\lambda$-values
  arbitrarily close to zero \cite{randommatrix3}, and verifies the well known Banks-Casher relation\footnote{One could also obtain a
  nonvanishing chiral condensate in the chiral limit if the spectral density has a contribution that behaves as
  $\rho\left(\lambda,m\right)_{\lambda\sim 0}\sim A\left(\frac{m}{\lambda}\right)^\alpha$
  with $0<\alpha<1$. Note however that this would give rise to a nonanalytic quark-mass contribution to the
  chiral condensate}.                                                                           

  \begin{equation}
    \lim_{\left(\lambda,m\right)\rightarrow\left(0^+,0^+\right)}\hskip0,1cm \rho\left(\lambda,m\right) = \rho\left(0,0\right) =
    \frac{\Sigma}{\pi}
\label{banks}                                                                                                                             
\end{equation}

  We want to note that although the chiral condensate is, in one-flavor $QCD$, an analytical function of the
  quark mass, the above discussion tells us that both, the two contributions to the scalar
condensate in (\ref{scalarc}) are nonanalytic functions of the quark mass, and their nonanalyticities
are exactly compensated in the sum \cite{randommatrix2}. In other words, the chiral and thermodynamic limits commute
on the sum (\ref{scalarc}), but not on each single addend.

\section{Spectral density in two-flavor $QCD$} \label{spectraltwo}

The G-W fermion action for the two-flavor model is

$$
S_F = \sum_{v, w} \bar\psi_u\left(v\right) D\left(v, w\right)\psi_u\left(w\right) +
m_u \sum_{v, w} \bar\psi_u\left(v\right) \left(1-\frac{1}{2}D
\left(v, w\right)\right)\psi_u\left(w\right) +
$$
\begin{equation}
\hskip0,4cm \sum_{v, w} \bar\psi_d\left(v\right) D\left(v, w\right)\psi_d\left(w\right) +
m_d \sum_{v, w} \bar\psi_d\left(v\right) \left(1-\frac{1}{2}D
\left(v, w\right)\right)\psi_d\left(w\right)
  \label{fa2f}
\end{equation}
where $m_u$ and $m_d$ are the up and down quark masses in lattice units. In the chiral limit action
(\ref{fa2f}) is invariant under the chiral rotations

\begin{equation}
  \begin{pmatrix}
    \psi_u\\
    \psi_d\\
  \end{pmatrix}
  \rightarrow e^{i\alpha\gamma_5M\left(I-\frac{1}{2}D\right)}
  \begin{pmatrix}
    \psi_u\\
    \psi_d\\
  \end{pmatrix}
  \hskip0,5cm
  \begin{pmatrix}
    \bar\psi_u & \bar\psi_d\\
  \end{pmatrix}
  \rightarrow
\begin{pmatrix}
    \bar\psi_u & \bar\psi_d\\
  \end{pmatrix}
  e^{i\alpha\gamma_5M\left(I-\frac{1}{2}D\right)}
  \label{chirals}
\end{equation}
where the $2\times2$ $M$-matrix can either be the identity or any Pauli matrix. These transformations
generate a $SU(2)_A\times U(1)_A$ symmetry group, but the Abelian $U(1)_A$ symmetry is anomalous, as in the
one-flavor model. 

The normalized up (or down) condensate, in the infinite volume limit, is

\begin{equation}
  \left<s_u\right> = \frac{1}{V} \left<\bar\psi_u \left(1-\frac{a}{2}D\right)\psi_u\right> = 
  -2m_u\int_0^2 \frac{\rho\left(\lambda,m_u,m_d\right)\left(1-\frac{1}{4}\lambda^2\right)}
      {m_u^2+ \left(1-\frac{1}{4}m_u^2\right)\lambda^2}d\lambda -
      \lim_{V\rightarrow\infty} \frac{1}{m_u} \left<\frac{n_+\left(U\right)+n_-\left(U\right)}{V}\right>,
  \label{sc2f}
\end{equation}
where we have, also in this case, explicitly written the contribution of the exact zero modes, and 
$\rho(\lambda,m_u,m_d)$ is the density of the absolute value of the complex conjugate eigenvalues
of the Dirac-Ginsparg-Wilson operator $D$

\begin{equation}
  \rho\left(\lambda,m_u,m_d\right) =
  \lim_{V\rightarrow\infty}
  \left<\frac{1}{V}\sum^{n.z.}_j\delta\left(\lambda-\lambda_j
  \left(U\right)\right)\right>
  \label{spectrald2f}
\end{equation}
with $0<\lambda_j(U)\le 2$.

For degenerate flavors, $m_u=m_d=m$, we can denote $\rho\left(\lambda,m_u,m_d\right)_{m_u=m_d}=\rho(\lambda,m)$, and we
write here some expressions that will be useful in what follows.

The scalar condensate is 

\begin{equation}
  \left<s\right> = \frac{1}{2}\left(\left<s_u\right>+\left<s_d\right>\right) = 
  -2m\int_0^2 \frac{\rho\left(\lambda,m\right)\left(1-\frac{1}{4}\lambda^2\right)}
      {m^2+ \left(1-\frac{1}{4}m^2\right)\lambda^2}d\lambda -
      \lim_{V\rightarrow\infty} \frac{1}{m} \left<\frac{n_+\left(U\right)+n_-\left(U\right)}{V}\right>
  \label{sc2fd}
\end{equation}

The pion susceptibility $\chi_\pi$ can be written, with the help of the Ward-Takahashi identity, as

\begin{equation}
  \chi_\pi = -\frac{\left<s_u\right>+\left<s_d\right>}{m} = 
  4\int_0^2 \frac{\rho\left(\lambda,m\right)\left(1-\frac{1}{4}\lambda^2\right)}
      {m^2+ \left(1-\frac{1}{4}m^2\right)\lambda^2}d\lambda +
      \lim_{V\rightarrow\infty} \frac{2}{m^2} \left<\frac{n_+\left(U\right)+n_-\left(U\right)}{V}\right>,
  \label{chipi}
\end{equation}
and the difference between the $\pi$ and $\delta$ susceptibilities is

\begin{equation}
  \chi_\pi - \chi_\delta = 
  8 m^2\int_0^2 \frac{\rho\left(\lambda,m\right)\left(1-\frac{1}{4}\lambda^2\right)^2}
      {\left(m^2+ \left(1-\frac{1}{4}m^2\right)\lambda^2\right)^2}d\lambda +
      \lim_{V\rightarrow\infty} \frac{4}{m^2} \left<\frac{n_+\left(U\right)+n_-\left(U\right)}{V}\right>
  \label{chipidelta}
\end{equation}

\subsection{Low temperature broken phase} \label{LT}

As far as non-Abelian $SU(2)_A$ chiral chiral symmetry is concerned, it is well known that $QCD$ spontaneously breaks
this symmetry in the low temperature phase, and in contrast to the single flavor model, 
the vacuum energy density and chiral condensates are not analytic functions of the quark masses, as a consequence of the
long-range order. The infinite volume limit and the chiral limit do not commute in this phase
and to analyze the structure of the vacuum, the chiral limit must be taken after the infinite volume limit.

The scalar condensate for degenerate flavors, which is an order parameter for $SU(2)_A$ chiral symmetry, is given by
Eq. (\ref{sc2fd}). The standard wisdom, as discussed in section \ref{spectral}, is that the contribution of
exact zero modes to the chiral condensate, second addend in
the right-hand side of (\ref{sc2fd}), should vanish in the thermodynamic limit because the density of zero modes
$\left<\frac{n_+\left(U\right)+n_-\left(U\right)}{V}\right>$ vanishes in this limit for every $m\ge 0$. Hence
spontaneous symmetry breaking is done by the Banks-Casher mechanism

\begin{equation}
  \Sigma_{2f} = -\lim_{m\rightarrow 0}\left<s\right> = 
  \lim_{m\rightarrow 0} 2m\int_0^2 \frac{\rho\left(\lambda,m\right)\left(1-\frac{1}{4}\lambda^2\right)}
      {m^2+ \left(1-\frac{1}{4}m^2\right)\lambda^2}d\lambda =
      \pi \rho(0,0)
  \label{sc2fdm0}
\end{equation}
with

\begin{equation}
  \rho\left(0,0\right) = 
  \lim_{\left(\lambda,m\right)\rightarrow\left(0^+,0\right)}\hskip0,1cm \rho\left(\lambda,m\right)
  \label{banks2}
\end{equation}

\subsection{High temperature symmetric phase} \label{HT}

It is also well known that at a critical temperature $T_c$, of the order of the pion mass,
two-flavor $QCD$ undergoes a phase transition to a
$SU(2)_A$ chiral symmetric phase. Moreover, as stated in the introduction of this article,
we argued in \cite{vic1,vic2,vic3}, using general properties of the $Q=0$
topological sector, that if the $U(1)_A$ axial symmetry remains effectively broken in the high temperature chiral
symmetric phase of $QCD$, the theory should exhibit a divergent correlation length
in the correlation function of the scalar condensate, as well as in other pseudoscalar correlation functions, in
the chiral limit.

A qualitative but powerful argument supporting this result is derived from the application of Landau's
theory of phase transitions to the expected phase diagram in the $Q=0$ topological sector \cite{vic1,vic3}.

\begin{figure}[h!]
  \centerline{\includegraphics[scale=1.03]{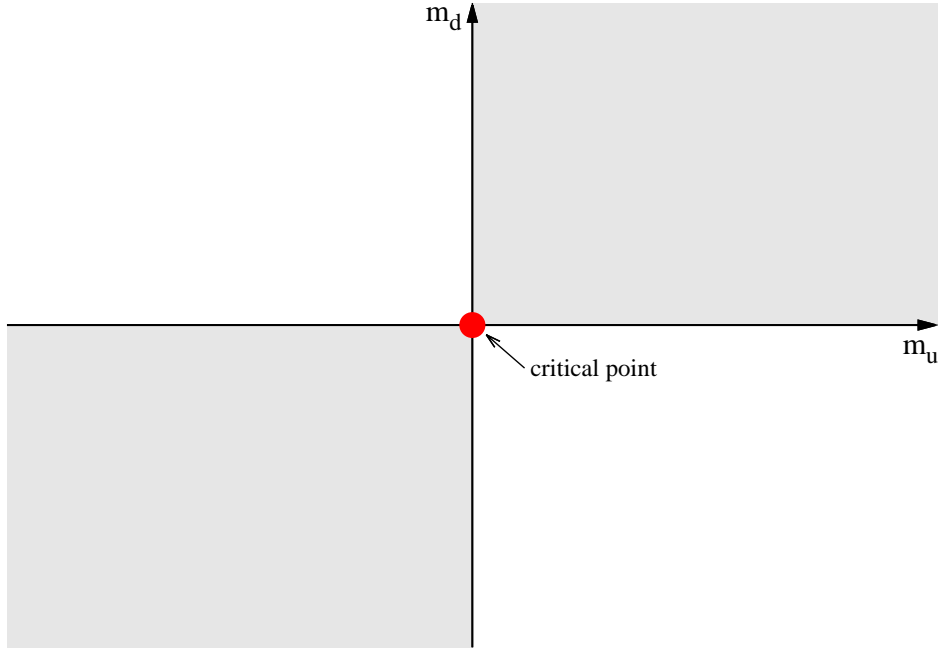}}
  \caption{Phase diagram of the two-flavor model in the $Q=0$ topological sector. The coordinate
axis in the $(m_u,m_d)$ plane are first order phase transition lines. The origin of coordinates
is the end point of all first order transition lines. The
vacuum energy density, its derivatives, and expectation values of local operators of
the two-flavor model at $\theta=0$ only agree with those of the $Q=0$ sector in the first
$(m_u>0,m_d>0)$ and third $(m_u<0,m_d<0)$ quadrants (the darkened areas).}
\end{figure}

We briefly summarize this argument here. Figure 1 is a schematic representation of the phase diagram
of the two-flavor model, in the
$Q=0$ topological sector, and in the $(m_u, m_d)$ plane. The two coordinate axis show first order
phase transition lines. All first order
transition lines end however at a common point, the origin of coordinates $m_u=m_d=0$,
where all condensates vanish because at this point we recover the $SU(2)_A$ chiral symmetry.
Notice that in the low temperature phase, where the non-Abelian chiral symmetry is spontaneously broken,
the phase diagram in the $(m_u, m_d)$ plane would be the same as that of Fig. 1
with the only exception that the origin of coordinates is not an endpoint.

Landau's theory of phase transitions predicts that the end point placed at the origin of
coordinates in the $(m_u, m_d)$ plane is a critical point, the scalar condensate should show
a nonanalytic dependence on the fermion masses, and the scalar susceptibility should diverge.
But since the vacuum energy
density in the $Q=0$ topological sector, and its fermion mass derivatives, matches the vacuum
energy density and fermion mass derivatives in the full theory,
and the same is true for the critical equation of state, Landau's theory
of phase transitions
predicts a non-analytic dependence of the flavor singlet scalar condensate on the
fermion mass, and a divergent correlation length in the chiral limit of our full theory, in which
we take into account the contribution of all topological sectors.

The two-flavor Schwinger model, analyzed by Coleman years ago \cite{coleman}, is a paradigmatic case of realization of this
scenario, as discussed in \cite{vic2,vic3}.

However, as far as QCD is concerned, some approaches, such as the DIGA or the quasi-instanton picture \cite{qip1,qip2},  
assume the validity of the perturbative
expansion of the vacuum energy density in powers of the quark masses. 

The Dilute Instanton Gas Approximation, 
trying to describe physics at very high temperature,
assumes that at temperatures much higher than $T_c$ Debye screening would only allow
instantons of very small radius to exist, and hence the $QCD$ vacuum energy density of noninteracting
instantons should not suffer from infrared singularities. The vacuum energy density can therefore be
expanded in powers of the up and down quark masses, and its $\theta$-dependent piece,
for degenerate quark masses, is

\begin{equation}
E_{\theta}^{DIGA} = -2 m^2 z_I\left(T\right) \cos\theta
  \label{diga}
\end{equation}

The quasi-instanton picture goes further, and assumes the validity of the perturbative expansion for any temperature greater than
$T_c$ \cite{qip1}.

We will assume in the rest of this section that, contrary to the prediction
of Landau's theory, the vacuum energy density can be expanded in powers of the quark masses, and therefore the chiral and thermodynamic
limits commute, like in the one-flavor model. Then we will show that, in such a case, either
the spectral density $\rho\left(\lambda,m\right)$ of the absolute value of the 
nonzero modes of the Dirac-Ginsparg-Wilson operator develops a $m^2\delta(\lambda)$ contribution in the
thermodynamic limit, or the theory becomes $\theta$-independent
to second order on quark masses.

The expansion of the vacuum energy density up to second order reads as follows

\begin{equation}
  E\left(m_u, m_d\right) = E\left(0, 0\right) -  
  \frac{1}{2} m_u^2 \chi_{s_{u, u}} -
  \frac{1}{2} m_d^2 \chi_{s_{d, d}} -
  m_u m_d \chi_{s_{u, d}} + \dots
\label{taylor2f}
\end{equation}
The linear terms in (\ref{taylor2f}) vanish because the $SU(2)_A$ symmetry is fulfilled
in the vacuum, and 
$\chi_{s_{u, u}}, \chi_{s_{d, d}}$ and $\chi_{s_{u, d}}$ are the scalar up, down and up-down
susceptibilities respectively

$$
\chi_{s_{u, u}} = V \left<s_u^2\right>_{m_u=m_d=0}
$$
$$
\chi_{s_{d, d}} = V \left<s_d^2\right>_{m_u=m_d=0}
$$  
\begin{equation}
\chi_{s_{u, d}} = V \left<s_u s_d\right>_{m_u=m_d=0}
\label{ududs}
\end{equation}
where $s_u$ and $s_d$ are the scalar up and down condensates (\ref{sapc}), 
normalized by the lattice volume $V=L_s^3\times L_t$, with $L_t=\frac{1}{T}$, the inverse temperature in
lattice units. The disconnected contributions are absent in (\ref{ududs}) because the $SU(2)_A$ chiral
symmetry constrains $\left<s_u\right>_{m_u=m_d=0}$ and $\left<s_d\right>_{m_u=m_d=0}$ to vanish, and 
$\chi_{s_{u, u}} = \chi_{s_{d, d}}$ because of flavor
symmetry.

The scalar up and down susceptibilities for massless fermions, $\chi_{s_{u, u}}$ and $\chi_{s_{d, d}}$,
get all their contribution from the $\nu=0$ topological sector,
while $\chi_{s_{u, d}}$ gets all its contribution from the $\nu= ^+_- 1$ topological sectors.
Hence the vacuum energy density (\ref{taylor2f}) in the presence of a $\theta$-vacuum term is,
up to second order,

\begin{equation}
  E\left(m_u, m_d, \theta\right) = E\left(0, 0\right) -  
  \frac{1}{2} \left(m_u^2+m_d^2\right) \chi_{s_{u, u}} -
  m_u m_d \cos\theta \chi_{s_{u, d}} + \dots
\label{taylor2ftheta}
\end{equation}
with

\begin{equation}
  \chi_{s_{u, u}} = \chi_{s_{d, d}} = 
  2\int_0^2 \frac{\rho\left(\lambda,0\right)\left(1-\frac{1}{4}\lambda^2\right)}
      {\lambda^2}d\lambda
\label{chium0}
\end{equation}
and there is no contribution from zero modes to these susceptibilities.

On the other hand, $\chi_{s_{u, d}}$ can be written as

$$
  \chi_{s_{u, d}} = \lim_{V\rightarrow\infty} \lim_{\left(m_u,m_d\right)\rightarrow \left(0,0\right)}
  \frac{1}{m_u m_d}\left<\frac{\left(n_+\left(U\right) + n_-\left(U\right)\right)^2}{V}\right> 
$$
  \begin{equation}
\hskip -2cm
  = \lim_{V\rightarrow\infty} \lim_{\left(m_u,m_d\right)\rightarrow \left(0,0\right)}
  \left(\frac{2}{V m_u m_d} \frac{Z_1}{Z_0}\right)
  \label{chiudm0}
\end{equation}

One can easily derive from Eq. (\ref{taylor2ftheta}) the following relations for the $\pi$ and $\pi-\delta$
susceptibilities in the chiral limit at $\theta=0$

\begin{equation}
  \chi_{\pi_{m_u=m_d=0}}=\chi_{\sigma_{m_u=m_d=0}}=2\chi_{s_{u, u}} + 2\chi_{s_{u, d}}
\label{chipim0}
\end{equation}

\begin{equation}
  \chi_{\pi_{m_u=m_d=0}}-\chi_{\delta_{m_u=m_d=0}}=4\chi_{s_{u, d}}
\label{chipideltam0}
\end{equation}

We will analyze here the case of degenerate flavors, $m_u=m_d=m$. Our goal in the calculation that follows is to find
general properties of the spectral density that give an effective breaking of the $U(1)_A$ axial symmetry and a vacuum
energy density that can be expressed as an even power series in the quark mass, and therefore 
do not give spontaneous breaking of the $SU(2)_A$ chiral symmetry.

We want to remark that, as far as the spectral density $\rho(\lambda,m)$ is concerned, the only moderate assumption in what
follows is that it is a continuous function of the fermion mass, at $m=0$, for each $\lambda>0$.

Since we are assuming the validity of the perturbative expansion of the free energy density (\ref{taylor2f}), the chiral
and thermodynamic limits commute. Hence we can also use Eqs. (\ref{chipi}) and (\ref{chipidelta}) 
for the computation of the chiral limit of the $\pi$ and $\pi-\delta$ susceptibilities. Moreover, the suppression of
paired zero modes at high temperature is even stronger than at low temperature. Therefore the zero modes contribution
to these susceptibilities (second addends in (\ref{chipi}) and (\ref{chipidelta})) should vanish in the thermodynamic
limit, as in the low temperature phase.
Then, 
if we take the massless limit of $\chi_\pi-\chi_\delta$, Eq. (\ref{chipidelta}), we can write

\begin{equation}
  \lim_{m\rightarrow 0} \chi_\pi-\chi_\delta = 
 \lim_{m\rightarrow 0}8 m^2\int_0^2 \frac{\rho\left(\lambda,m\right)\left(1-\frac{1}{4}\lambda^2\right)^2}
     {\left(m^2+ \left(1-\frac{1}{4}m^2\right)\lambda^2\right)^2}d\lambda =
     \lim_{m\rightarrow 0}8 m^2\int_0^{\epsilon} \frac{\rho\left(\lambda,m\right)\left(1-\frac{1}{4}\lambda^2\right)^2}
     {\left(m^2+ \left(1-\frac{1}{4}m^2\right)\lambda^2\right)^2}d\lambda
\label{epsilon1}
\end{equation}
for every $\epsilon>0$, and taking into account Eq. (\ref{chipideltam0}) we get

\begin{equation}
  \lim_{m\rightarrow 0}2 m^2\int_0^{\epsilon} \frac{\rho\left(\lambda,m\right)\left(1-\frac{1}{4}\lambda^2\right)^2}
     {\left(m^2+ \left(1-\frac{1}{4}m^2\right)\lambda^2\right)^2}d\lambda  = \chi_{s_{u, d}}
  \label{chisud1}
\end{equation}

By taking on the other hand the massless limit of the pion susceptibility $\chi_\pi$, Eq. (\ref{chipi}), we can write

$$
  \lim_{m\rightarrow 0}\chi_\pi = 
  \lim_{m\rightarrow 0}4\int_0^2 \frac{\rho\left(\lambda,m\right)\left(1-\frac{1}{4}\lambda^2\right)}
      {m^2+ \left(1-\frac{1}{4}m^2\right)\lambda^2}d\lambda
$$
$$
      = \lim_{\epsilon\rightarrow 0}\lim_{m\rightarrow 0}
      4\int_0^{\epsilon} \frac{\rho\left(\lambda,m\right)\left(1-\frac{1}{4}\lambda^2\right)}
      {m^2+ \left(1-\frac{1}{4}m^2\right)\lambda^2}d\lambda +
      \lim_{\epsilon\rightarrow 0}4\int_{\epsilon}^2 \frac{\rho\left(\lambda,0\right)\left(1-\frac{1}{4}\lambda^2\right)}
      {\lambda^2}d\lambda
$$
      \begin{equation}
      = \lim_{\epsilon\rightarrow 0} \lim_{m\rightarrow 0}
      4\int_0^{\epsilon} \frac{\rho\left(\lambda,m\right)\left(1-\frac{1}{4}\lambda^2\right)}
      {m^2+ \left(1-\frac{1}{4}m^2\right)\lambda^2}d\lambda +
      4\int_0^2 \frac{\rho\left(\lambda,0\right)\left(1-\frac{1}{4}\lambda^2\right)}
      {\lambda^2}d\lambda
  \label{epsilon2}
\end{equation}
      \footnote{We have implicitly excluded in Eq. (\ref{epsilon2}) an unphysical behavior
      $\rho(\lambda,0)\sim \lambda^2\delta(\lambda)$ around $\lambda=0$ since it would give rise to an infrared divergence in the
      higher order perturbative expansion (\ref{taylor2f}).}, 
and taking into account Eqs. (\ref{chium0}) and (\ref{chipim0}) we get

\begin{equation}
\lim_{\epsilon\rightarrow 0} \lim_{m\rightarrow 0}
      2\int_0^{\epsilon} \frac{\rho\left(\lambda,m\right)\left(1-\frac{1}{4}\lambda^2\right)}
      {m^2+ \left(1-\frac{1}{4}m^2\right)\lambda^2}d\lambda = \chi_{s_{u, d}}
  \label{chisud2}
\end{equation}

Eqs. (\ref{chisud1}) and (\ref{chisud2}) allow us to write

\begin{equation}
  \lim_{\epsilon\rightarrow 0} \lim_{m\rightarrow 0}
  \int_0^{\epsilon} \frac{\rho\left(\lambda,m\right)\left(1-\frac{1}{4}\lambda^2
    \right)}{m^2+ \left(1-\frac{1}{4}m^2\right)\lambda^2}d\lambda =
    \lim_{m\rightarrow 0} m^2\int_0^{\epsilon} \frac{\rho\left(\lambda,m\right)\left(1-\frac{1}{4}\lambda^2\right)^2}
        {\left(m^2+ \left(1-\frac{1}{4}m^2\right)\lambda^2\right)^2}d\lambda =
        \frac{1}{2}\chi_{s_{u, d}}.
  \label{vaya}
\end{equation}
It is easy to show that this equation is equivalent to

\begin{equation}
  \lim_{\epsilon\rightarrow 0} \lim_{m\rightarrow 0}
  \int_0^{\epsilon} \frac{\rho\left(\lambda,m\right)}{m^2+
    \left(1-\frac{1}{4}m^2\right)\lambda^2}d\lambda =
  \lim_{m\rightarrow 0} m^2\int_0^{\epsilon} \frac{\rho\left(\lambda,m\right)}
      {\left(m^2+ \left(1-\frac{1}{4}m^2\right)\lambda^2\right)^2}d\lambda =
      \frac{1}{2}\chi_{s_{u, d}}
  \label{vaya2}
\end{equation}
and making the subtraction we get

\begin{equation}
  \lim_{\epsilon\rightarrow 0} \lim_{m\rightarrow 0}
  \int_0^{\epsilon} \frac{\rho\left(\lambda,m\right)\left(1-\frac{1}{4}m^2\right)\lambda^2}{\left(m^2+
    \left(1-\frac{1}{4}m^2\right)\lambda^2\right)^2}d\lambda = 0
  \label{vaya3}
\end{equation}
that is to say

\begin{equation}
  \lim_{\epsilon\rightarrow 0} \lim_{m\rightarrow 0}
  \int_0^{\epsilon} \frac{\rho\left(\lambda,m\right)\lambda^2}{\left(m^2+
    \left(1-\frac{1}{4}m^2\right)\lambda^2\right)^2}d\lambda = 0
  \label{vaya4}
\end{equation}

The last point of our argument is to show that compatibility of Eqs. (\ref{vaya2}) and (\ref{vaya4})
requires that the spectral density $\rho(\lambda,m)$ have a singular contribution

\begin{equation}
  \rho_{sing}\left(\lambda,m\right) = \chi_{s_{u, d}} m^2 \delta\left(\lambda\right)
  \label{vaya5}
\end{equation}
in addition to other terms that contribute neither to (\ref{vaya2}) nor to
(\ref{vaya4}).

The first step in proving Eq. (\ref{vaya5}) is to write
Eqs. (\ref{vaya2}) and (\ref{vaya4}) in a slightly modified mode

$$
\lim_{m\rightarrow 0} \int_0^{\epsilon} \frac{m^2\rho\left(\lambda,m\right)}
    {\left(m^2+ \left(1-\frac{1}{4}m^2\right)\lambda^2\right)^2}d\lambda =
    \lim_{m\rightarrow 0} \int_0^{m} \frac{m^2\rho\left(\lambda,m\right)}
        {\left(m^2+ \left(1-\frac{1}{4}m^2\right)\lambda^2\right)^2}d\lambda +
$$
\begin{equation}
\hskip6cm \lim_{m\rightarrow 0} \int_m^{\epsilon} \frac{m^2\rho\left(\lambda,m\right)}
                    {\left(m^2+ \left(1-\frac{1}{4}m^2\right)\lambda^2\right)^2}d\lambda
                    = \frac{1}{2}\chi_{s_{u, d}}
\label{vaya7}
\end{equation}
and

$$
\lim_{\epsilon\rightarrow 0} \lim_{m\rightarrow 0}
\int_0^{\epsilon} \frac{\lambda^2\rho\left(\lambda,m\right)}{\left(m^2+
  \left(1-\frac{1}{4}m^2\right)\lambda^2\right)^2}d\lambda =
\lim_{m\rightarrow 0}
\int_0^{m} \frac{\lambda^2\rho\left(\lambda,m\right)}{\left(m^2+
  \left(1-\frac{1}{4}m^2\right)\lambda^2\right)^2}d\lambda +
$$
\begin{equation}
  \hskip5.5cm \lim_{\epsilon\rightarrow 0} \lim_{m\rightarrow 0}
  \int_m^{\epsilon} \frac{\lambda^2\rho\left(\lambda,m\right)}{\left(m^2+
    \left(1-\frac{1}{4}m^2\right)\lambda^2\right)^2}d\lambda
  = 0
  \label{vaya8}
\end{equation}
Since all the above integrals are non-negative, the two terms in the right-hand side of equation (\ref{vaya8})
must cancel independently. Moreover the second integral in the right-hand side of (\ref{vaya7}) is upper-bounded by
the second integral in the right-hand side of (\ref{vaya8}), and hence the following equations have to be fulfilled

\begin{equation}
  \lim_{m\rightarrow 0} \int_0^{m} \frac{m^2\rho\left(\lambda,m\right)}
      {\left(m^2+ \left(1-\frac{1}{4}m^2\right)\lambda^2\right)^2}d\lambda =
      \frac{1}{2}\chi_{s_{u, d}}
      \label{vaya9}
\end{equation}

\begin{equation}
\lim_{m\rightarrow 0}
\int_0^{m} \frac{\lambda^2\rho\left(\lambda,m\right)}{\left(m^2+
  \left(1-\frac{1}{4}m^2\right)\lambda^2\right)^2}d\lambda = 0
\label{vaya10}
\end{equation}

Now, if we perform the variable change $\lambda = m\mu$ in the previous integrals we get

\begin{equation}
\lim_{m\rightarrow 0}\int_0^{1} \frac{\rho\left(\mu m,m\right)}
    {m\left(1+ \left(1-\frac{1}{4}m^2\right)\mu^2\right)^2}d\mu = \frac{1}{2}\chi_{s_{u, d}}
    \label{vaya11}
\end{equation}

\begin{equation}
  \lim_{m\rightarrow 0}\int_0^{1} \frac{\mu^2\rho\left(\mu m,m\right)}{m\left(1+
    \left(1-\frac{1}{4}m^2\right)\mu^2\right)^2}d\mu = 0
  \label{vaya12}
\end{equation}
and defining the following normalized probability distribution function

\begin{equation}
  P\left(\mu,m\right) =
  \left(\int_0^{1} \frac{\rho\left(\mu m,m\right)}
       {m\left(1+ \left(1-\frac{1}{4}m^2\right)\mu^2\right)^2}d\mu\right)^{-1}
       \frac{\rho\left(\mu m,m\right)}
            {m\left(1+ \left(1-\frac{1}{4}m^2\right)\mu^2\right)^2}
            \label{vaya13}
\end{equation}
one can write

\begin{equation}
  \lim_{m\rightarrow 0}
  \int_0^{1} \mu^2 P\left(\mu,m\right) d\mu = 0
  \label{vaya14}
\end{equation}
and in general

\begin{equation}
  \lim_{m\rightarrow 0}
  \int_0^{1} \mu^\alpha P\left(\mu,m\right) d\mu = 0 \hskip 1cm \forall \alpha> 0
  \label{vaya14b}
\end{equation}

Equation (\ref{vaya14b}) implies that the normalized probability distribution function
$P\left(\mu,m\right)$ approaches a Dirac delta, $2\delta(\mu)$, as m approaches zero, and hence we can write

\begin{equation}
  \lim_{m\rightarrow 0}\frac{\rho\left(\mu m,m\right)}
      {m\left(1+ \left(1-\frac{1}{4}m^2\right)\mu^2\right)^2} =
      \chi_{s_{u, d}} \delta\left(\mu\right)
      \label{vaya15}
\end{equation}
which gives rise to the following singular contribution to the spectral density $\rho(\lambda,m)$

\begin{equation}
  \rho_{sing}\left(\lambda,m\right) = \frac{\chi_{s_{u, d}}}{m^2}\left(m^2 + \left(1-\frac{1}{4}m^2\right)
  \lambda^2\right)^2
  \delta\left(\lambda\right) = \chi_{s_{u, d}} m^2 \delta\left(\lambda\right)
  \label{vaya16}
\end{equation}

Therefore we can conclude that the only way to consistently derive a nonzero value of $\chi_{s_{u, d}}$ from an
analytic free energy density (\ref{taylor2f}) is that the spectral density $\rho\left(\lambda,m\right)$ of the absolute value of the 
nonzero modes of the Dirac-Ginsparg-Wilson operator develops a $\delta(\lambda)$ function in the thermodynamic limit,
which implies a nonzero density of zero modes in this limit. 

Equation (\ref{vaya16}) is expected to hold in the DIGA, which assumes that the
density of random instantons is $O(m^2)$ \cite{shur}. This finite density of random instantons implies the existence of
near zero-modes, whose contribution to the spectral density $\rho(\lambda,m)$ should be well approximated, at temperatures
much higher than $T_c$, by \cite{shur,rajan}:

\begin{equation}
  \rho\left(\lambda,m\right) \approx z_I\left(T\right) m^2 \delta\left(\lambda\right)
  \label{vaya17}
\end{equation}

The use of the delta function in (\ref{vaya17}) requires the implicit assumption that the small splitting from zero
of the near-zero modes, produced by the interactions between instantons and anti-instantons in the dilute gas, can be
neglected.
Nevertheless at lower temperatures, and especially at temperatures
close to $T_c$, the interaction between instantons should become non-negligible, and the semiclassical perturbative treatment of widely 
separated instantons and anti-instantons should no longer be valid. Therefore we expect that Eq. (\ref{vaya17}) does not hold 
at these temperatures. But we have shown that, if Eq. (\ref{vaya17}) does not hold, the free energy
density cannot be an analytic function of the quark mass, and therefore the Landau's theory prediction,
discussed at the beginning of this section, becomes the most plausible scenario.

\section{Conclusions} \label{con}

Using general properties of the $Q=0$ topological sector we argued in
Refs. \cite{vic1,vic2,vic3} that 
a vectorlike theory with chiral $U(1)_A$ anomaly, in which the $U(1)_A$ axial symmetry remains effectively broken, and 
where the chiral condensate vanishes in the chiral limit, because of a not spontaneously broken
non-Abelian chiral symmetry,
should exhibit a divergent correlation length in the correlation function of the scalar condensate in
the chiral limit. In such a case also some pseudoscalar correlation functions, associated to what would be the
Nambu-Goldstone bosons if the non-Abelian chiral symmetry were spontaneously broken,
should exhibit a divergent correlation length. The two-flavor Schwinger model, analyzed by Coleman \cite{coleman}, is a paradigmatic
example of realization of this scenario, as discussed in \cite{vic2,vic3}.

$QCD$ at $T>T_c$ satisfies all the above conditions, and it is also expected that the 
$U(1)_A$ axial symmetry remains effectively broken in its high temperature phase.
Therefore we would expect, based on the results of Refs. \cite{vic1,vic2,vic3},
a nonanalyticity in the quark mass dependence of the free energy density, in contrast to the DIGA prediction.

We have investigated in this work whether the aforementioned results can also be
reproduced
making only use of standard properties of the spectral density of the Dirac operator,
without having to resort to general properties of the $Q=0$ topological sector.
To this end we have assumed that the free energy density of $QCD$ with two degenerate
flavors is, in the high temperature phase, an analytic function of the quark mass,
and have shown that, in such a case,
the only way to derive a nontrivial $\theta$-dependence of an analytical free energy
density is that the spectral density, $\rho\left(\lambda,m\right)$, of the absolute value of the
nonzero modes of the Dirac-Ginsparg-Wilson operator develops a $m^2\delta(\lambda)$
function in the thermodynamic limit. We want to stress that this kind of behavior
of $\rho\left(\lambda,m\right)$ is crucial to get simultaneously analyticity and
nontrivial $\theta$-dependence. Any other kind of behavior of
$\rho\left(\lambda,m\right)$ around $\lambda=0$ would be incompatible with an analytical
free energy density or would lead to a $\theta$-independent theory, at least to second
order in the quark mass. We also want to point out that, as far as the spectral density $\rho(\lambda,m)$ is concerned, the
only moderate assumption we have made is that it is a continuous function of the fermion mass, at $m=0$, for each $\lambda>0$.

A $m^2\delta(\lambda)$ contribution to the spectral density $\rho(\lambda,m)$
is expected in the DIGA, where interactions between instantons are
fully neglected. This approximation may be reliable at very high temperatures, but 
at temperatures close to $T_c$ 
the interaction between instantons should become non-negligible, and 
the splitting from zero of the near-zero modes, which has been neglected in the DIGA,
should be taken into account. Therefore we can expect that the
$m^2\delta(\lambda)$ contribution to the spectral density is no longer correct. But we have shown that 
in such a case the free energy density can no longer be an analytic function of the quark mass,
a result that agrees with the predictions of Refs. \cite{vic1,vic2,vic3}. Therefore we conclude that the Landau's theory approach,
summarized at the beginning of section \ref{HT}, becomes the most plausible scenario to explain the origin of the nonanalyticity
of the free energy density.

\section{Acknowledgments} \label{aknow}

The author thanks Jacobus Verbaarschot for useful comments, and	Matteo Giordano	for many interesting discussions as well as for a
critical reading of the manuscript. This work was funded by MCIN/AEI/10.13039/501100011033 and “ERDF A way of making Europe” under
Grant No. PGC2022-126078NB-C21, and DGA-FSE under Grant No. 2020-E21-17R.

\vfill
\eject

\vfill
\eject

\end{document}